# Physics-constrained intraventricular vector flow mapping by color Doppler

**Florian Vixège[1], Alain Berod[2], Yunyun Sun[1], Simon Mendez[2], Olivier Bernard[1], Nicolas Ducros[1], Pierre-Yves Courand[1,3], Franck Nicoud[2]**, and **Damien Garcia[1]**

[1] CREATIS UMR 5220, U1294, University Lyon 1, INSA Lyon, France

[2] IMAG UMR 5149, University of Montpellier, France

[3] Department of echocardiography, Croix-Rousse Hospital, Lyon, France

E-mails: Garcia.Damien@gmail.com; Damien.Garcia@inserm.fr; Florian.Vixege@creatis.insa-lyon.fr

**Abstract.** Color Doppler by transthoracic echocardiography creates two-dimensional fan-shaped maps of blood velocities in the cardiac cavities. It is a one-component velocimetric technique since it only returns the velocity components parallel to the ultrasound beams. *Intraventricular vector flow mapping* (*i*VFM) is a method to recover the blood velocity vectors from the Doppler scalar fields in an echocardiographic three-chamber view. We improved our *i*VFM numerical scheme by imposing physical constraints. The *i*VFM consisted in minimizing regularized Doppler residuals subject to the condition that two fluid-dynamics constraints were satisfied, namely planar mass conservation, and free-slip boundary conditions. The optimization problem was solved by using the Lagrange multiplier method. A finite-difference discretization of the optimization problem, written in the polar coordinate system centered on the cardiac ultrasound probe, led to a sparse linear system. The single regularization parameter was determined automatically for non-supervision considerations. The physics-constrained method was validated using realistic intracardiac flow data from a patient-specific CFD (computational fluid dynamics) model. The numerical evaluations showed that the *i*VFM-derived velocity vectors were in very good agreement with the CFD-based original velocities, with relative errors ranged between 0.3 and 12%. We calculated two macroscopic measures of flow in the cardiac region of interest, the mean vorticity and mean stream function, and observed an excellent concordance between physics-constrained *i*VFM and CFD. The capability of physics-constrained *i*VFM was finally tested with *in vivo* color Doppler data acquired in patients routinely examined in the echocardiographic laboratory. The vortex that forms during the rapid filling was deciphered. The physics-constrained *i*VFM algorithm is ready for pilot clinical studies and is expected to have a significant clinical impact on the assessment of diastolic function.

# 1. Introduction

Its accessibility, and its ability to provide noninvasive information in real time, make echocardiography the standard technique for the evaluation of cardiac function. Echocardiographic assessment of left ventricular diastolic function includes measurements of venous and pulmonary flows, as well as the examination of transmitral blood velocities and mitral annulus velocities. These parameters describe different characteristics of left ventricular filling, and their analysis can help assess diastole. However, the diagnosis of diastolic dysfunction is often imprecise because the recommended echocardiographic indices may present discordant results. A thorough analysis of intraventricular flow could change this situation. To date, only local measurements of blood velocity, using continuous or pulsed wave spectral Doppler, are used for clinical diagnostic purposes. Although it is possible to obtain two-dimensional Doppler mapping, color Doppler is primarily qualitative in a clinical context. No quantitative color Doppler method has yet proven its routine clinical value at the bedside, with the exception of the proximal isovelocity surface area (PISA) method for grading mitral regurgitation, a technique subject to practical pitfalls (Grayburn and Thomas, 2021). Another quantitative technique based on M-mode color Doppler, which estimates intraventricular pressure differences (Yotti et al., 2005; Hodzic et al., 2020), may be of diagnostic value, although no clinical studies have yet really provided evidence for this.

The clinical context of the present study is two-dimensional color Doppler imaging in the left ventricle, with the planned objective of deciphering blood flow during cardiac filling (diastole) and quantifying the vortical flow structures. During diastole, the mitral valve forces the left intraventricular flow to create a vortex, i.e. a swirling mass of blood. This vortex directs blood to the left ventricular outflow tract (i.e., the outflow towards the aorta). In healthy subjects, it facilitates the transition from filling to ejection. When filling is impaired (diastolic dysfunction), there is a change in blood flow, with a significant impact on this intraventricular vortex. According to recent literature, it is manifest that the properties of the vortex are related to filling function (Bermejo et al., 2015; Arvidsson et al., 2016). There are a limited number of clinical imaging tools for the non-invasive analysis of intracardiac blood flow. Phase-contrast cardiac magnetic resonance (PC-CMR) can provide a time-resolved volumetric characterization of blood flow in the left ventricle at a sufficiently precise spatial resolution. CMR velocimetry, however, is not implemented in a routine clinical setting due to its limited accessibility and long acquisition time. Echo-PIV (echographic particle image velocimetry) yields an efficient echographic tool for intra-ventricular flow mapping. This technique, applied to contrast-enhanced echo images, can track ultrasound speckles to estimate blood motion within image planes. It requires a continuous intravenous injection of contrast agent to reach an image quality suitable for motion tracking (Garcia et al., 2017). Although no major side effect has been noticed, this procedure is time- and staff-consuming. Echo-PIV thus cannot be recommended for routine clinical practice. To address this issue, a contrast-free high-frame-rate procedure called "blood speckle imaging" has been introduced in GE clinical scanners to track the native speckles of blood in pediatric or transesophageal ultrasound imaging. The team behind this approach has evaluated it clinically in the scope of pediatric echocardiography (Fadnes et al., 2014; Nyrnes et al., 2020).

Another imaging modality for intraventricular vector flow imaging is $i$VFM (intraventricular vector flow mapping). The $i$VFM technique derives velocity vectors from conventional color Doppler. Color Doppler is a planar one-component velocimetric method; it returns velocity components parallel to the ultrasound beams. The objective of $i$VFM is to recover two-component planar information from these incomplete flow data. The concept of retrieving two-dimensional vector maps from color Doppler velocities was first introduced by (Ohtsuki and Tanaka, 2006), then reported concomitantly in (Garcia et al., 2010) and (Uejima et al., 2010). The $i$VFM method proposed by Garcia *et al.* consists in computing the transverse (angular) velocity components from the Doppler (radial) velocities by integrating the 2-D continuity equation across the ultrasound beamlines, i.e. along the isoradial lines. This $i$VFM flow-vector modality has been implemented in FUJIFILM Healthcare (formerly Hitachi) ultrasound machines (Tanaka et al., 2015) and has been the tool of recent clinical studies to investigate intraventricular flows in some cardiomyopathies (Ro et al., 2014; Stugaard et al., 2015). The first published $i$VFM technique (Garcia et al., 2010), which is used in

FUJIFILM Healthcare scanners, examines each isoradial line independently, thus generating vector discontinuities along the radial direction that must be post-processed by smoothing. Incorrect apical alignments can lead to significant inconsistencies. To overcome the shortcomings of this line-by-line strategy, we subsequently proposed a global minimization method (Assi et al., 2017). In a few words, we minimized a least-squares cost function involving four terms related to the input Doppler data, the conservation equation, the boundary conditions, and a smoothing regularization term. The cost function includes three regularizing scalars. We determined these parameters automatically through an *L*-hypercurve (Belge et al., 2002) to make the algorithm operator-independent. The inclusion of these three parameters makes the problem somewhat burdensome.

To improve the numerical implementation of *i*VFM and reduce to a single regularization parameter, we now propose an optimization problem that imposes two physics-based constraints. The *i*VFM problem is solved under the condition that two fluid-dynamics constraints are satisfied: mass conservation, i.e. free-divergence velocity field, and free-slip boundary conditions. Alike the previous version, the minimization problem is discretized with finite differences. Unlike the previous version, the argument that minimizes the cost function is determined, subject to equality of constraints, by the method of Lagrange multipliers. Consistent with (Assi et al., 2017), we evaluated the performance of the physics-constrained *i*VFM modality in a patient-specific computational fluid dynamics (CFD) cardiac model. We then tested it in a few patients to investigate its clinical feasibility.

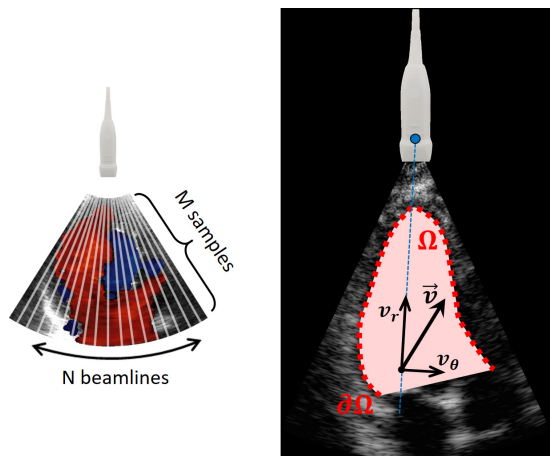

**Figure 1.** The physics-constrained *i*VFM algorithm is implemented in the polar coordinate system associated with the color-Doppler sector. Physics-constrained *i*VFM returns a 2-D velocity field $\vec{v}(r, \theta)$ from the Doppler components by minimizing a cost function subject to constraint equalities. $\Omega$ represents the domain of interest, and $\partial\Omega$ is its boundary.

## 2. Methods

### 2.1. Physics-constrained iVFM for vector flow reconstruction

Figure 1 illustrates a three-chamber (apical long-axis) view from transthoracic Doppler echocardiography. We consider the polar coordinate system $\{r, \theta\}$ whose pole is the center of the scan sector. In conventional cardiac ultrasound imaging, the successive ultrasound beamlines that form the image have a radial direction (Fig. 1, left). Color Doppler returns the blood velocity components parallel to these scanlines (Fig. 1, right) with additive noise. By convention, the Doppler velocities $u_{\mathrm{D}}$ are positive when the blood flows towards the ultrasound probe. In the following, bold notation represents vector (bold lowercase letters) or matrix (bold uppercase letters). As in (Assi et al., 2017), we define the velocity $v_{\mathrm{D}} = -u_{\mathrm{D}}$ to ensure sign compatibility between $v_{\mathrm{D}}$ and the radial components $v_r$ of the actual velocity field $\boldsymbol{v}$. Using this notation, color Doppler provides partial velocity information:

$$v_{\mathrm{D}}(r,\theta) = \boldsymbol{v}(r,\theta) \cdot \boldsymbol{e_r} + \eta(r,\theta) = v_r(r,\theta) + \eta(r,\theta), \tag{1}$$

where $\boldsymbol{e_r}$ is the unit radial vector, and $\eta$ is the Doppler noise. From this scalar noisy field, we seek to estimate the radial and angular components $\{v_r, v_\theta\}$ of the actual blood velocity field. Let $\{\hat{v}_r, \hat{v}_\theta\}$ stand for the components of the estimated velocity field $\hat{\boldsymbol{v}}(r,\theta)$. Let $\Omega$ be the domain of interest (Figure 1) that represents the left intraventricular cavity, with its endocardial boundary $\partial\Omega$. In the physics-constrained $i$VFM (Figure 2), the velocity field estimation problem is written as a minimization problem subject to two equality constraints. We aim for the radial velocities to be closely related to the input Doppler data, provided that the two-dimensional velocity vector field satisfies two physics restrictions. Mathematically, we write the $i$VFM problem as follows:

$$\{\hat{v}_r, \hat{v}_\theta\} = \arg \min_{(v_r, v_\theta)} \underbrace{\left\{ \int_\Omega \omega\, (v_r - v_{\mathrm{D}})^2\, \mathrm{d}\Omega \right\}}_{\substack{\text{closely match the} \\ \text{Doppler data}}} \tag{2}$$

subject to:

1. $r\,\mathrm{div}(\hat{\boldsymbol{v}}) = r\frac{\partial \hat{v}_r}{\partial r} + \hat{v}_r + \frac{\partial \hat{v}_\theta}{\partial \theta} = 0 \quad \text{on } \Omega,$
2. $(\hat{\boldsymbol{v}} - \boldsymbol{v_W}) \cdot \boldsymbol{n_W} = (\hat{v}_r - v_{Wr})\, n_{Wr} + (\hat{v}_\theta - v_{W\theta})\, n_{W\theta} = 0 \quad \text{on } \partial\Omega.$

The term $\omega$ stands for weights that are allocated to *in vivo* Doppler data (more details later). The subscript $(W)$ refers to the inner wall (endocardium). The vector $\boldsymbol{n_W} = \{n_{Wr},\, n_{W\theta}\}$ is a unit vector perpendicular (normal) to the endocardial wall. The vector $\boldsymbol{v_W} = \{v_{Wr},\, v_{W\theta}\}$ is a velocity vector of the endocardial wall.

1) The first equality constraint ensures that a divergence-free velocity vector field is returned. Since we work in two dimensions, this mass conservation implies that the out-of-plane components are zero. As shown in (Garcia et al., 2010), the 2-D divergence-free assumption is acceptable on the plane corresponding to the three-chamber apical long-axis view (Figure 1).

2) The second equality constraint is related to free-slip conditions on the endocardial wall boundary $\partial\Omega$. The free-slip condition assumes that there are no viscous effects at the wall. This condition is appropriate because the spatial resolution of color Doppler is too low to capture the boundary layer.

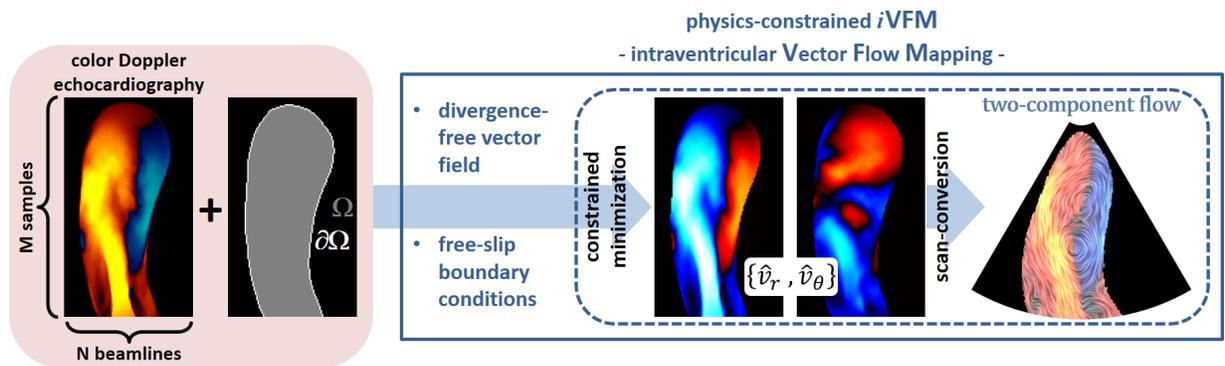

**Figure 2.** The physics-constrained $i$VFM algorithm is written as a minimization problem constrained by two physical properties: mass conservation and no penetration boundary conditions. It works with color Doppler velocities before scan conversion, in a polar coordinate system (leftmost images). $\Omega$ represents the domain of interest (left ventricular cavity), and $\partial\Omega$ is its boundary (endocardial wall). Figure adapted from (Assi et al., 2017).

We computed the solution of the constrained minimization problem (2) over the polar grid of the color Doppler (before scan conversion), which is an evenly spaced grid with constant radial and angular steps ($h_r$ and $h_\theta$). The differential operators were replaced by their discrete counterparts using second-order central finite differences with three-point stencils. We introduced the matrices described in Table 1, all of size ($M \times N$), where $N$ is the number of beamlines and $M$ is the number of samples per beamline (Figure 2). Table 1 also reports the corresponding column vectors of size ($MN \times 1$) after their vectorization. The Hadamard (entrywise) and Kronecker products are noted $\circ$ and $\otimes$, respectively. For a vector $\boldsymbol{a}$, the entrywise square is noted $\boldsymbol{a}^{\circ 2} = \boldsymbol{a} \circ \boldsymbol{a}$. Similarly, its entrywise root mean square is noted $\boldsymbol{a}^{\circ \frac{1}{2}}$. The operator diag($\boldsymbol{a}$) denotes a square diagonal matrix with the elements of the vector $\boldsymbol{a}$ on the main diagonal. By using the column arrays and matrices defined in Table 1, a discretized form of the constrained minimization problem (2) can be written as:

$$\hat{\boldsymbol{v}} = \{\hat{v}_r, \hat{v}_\theta\} = \arg\min_{\boldsymbol{v}} \{ \left( Q_0 \boldsymbol{v} - \text{diag}\left( \boldsymbol{\omega}^{\circ \frac{1}{2}} \circ \boldsymbol{\delta} \right) \boldsymbol{v}_D \right)^{\text{T}} \left( Q_0 \boldsymbol{v} - \text{diag}\left( \boldsymbol{\omega}^{\circ \frac{1}{2}} \circ \boldsymbol{\delta} \right) \boldsymbol{v}_D \right) \}$$

subject to:
$$\begin{cases} Q_1 \boldsymbol{v} = \mathbb{O}_{MN}, \\ Q_2 (\boldsymbol{v} - \boldsymbol{v}_W) = \mathbb{O}_{MN}, \end{cases}$$

(3)

where $\boldsymbol{Q_0}, \boldsymbol{Q_1}, \boldsymbol{Q_2}$ are three sparse matrices of size ($MN \times 2MN$). They are similar to those introduced in (Assi et al., 2017) and are given by:

$$Q_0 = \left[ \text{diag}\left( \boldsymbol{\omega}^{\circ \frac{1}{2}} \circ \boldsymbol{\delta} \right) \quad O_{MN} \right],$$
$$Q_1 = \left[ \frac{1}{h_r} \text{diag}(\boldsymbol{\delta} \circ \boldsymbol{r}) \left( I_N \otimes \dot{D}_M \right) + \text{diag}(\boldsymbol{\delta}) \quad , \quad \frac{1}{h_\theta} \text{diag}(\boldsymbol{\delta}) \left( \dot{D}_N \otimes I_M \right) \right],$$
$$Q_2 = \left[ \text{diag}(\boldsymbol{n_r}) \quad \text{diag}(\boldsymbol{n_\theta}) \right].$$

(4)

The Lagrangian function of the constrained minimization problem (3) is given by:

$$\mathcal{L}(\boldsymbol{v}, \boldsymbol{\lambda_1}, \boldsymbol{\lambda_2}) = \left( Q_0 \boldsymbol{v} - \boldsymbol{\omega}^{\circ \frac{1}{2}} \circ \boldsymbol{\delta} \circ \boldsymbol{v}_D \right)^{\text{T}} \left( Q_0 \boldsymbol{v} - \boldsymbol{\omega}^{\circ \frac{1}{2}} \circ \boldsymbol{\delta} \circ \boldsymbol{v}_D \right) + \boldsymbol{\lambda_1^{\text{T}}} Q_1 \boldsymbol{v} + \boldsymbol{\lambda_2^{\text{T}}} Q_2 (\boldsymbol{v} - \boldsymbol{v_w}).$$

(5)

Solving $\nabla_{\boldsymbol{v}, \boldsymbol{\lambda_1}, \boldsymbol{\lambda_2}} \mathcal{L}(\boldsymbol{v}, \boldsymbol{\lambda_1}, \boldsymbol{\lambda_2}) = 0$ leads to the linear system that contains the solution of the constrained minimization problem:

$$\underbrace{\begin{bmatrix} 2 \, Q_0^{\text{T}} Q_0 & Q_1^{\text{T}} & Q_2^{\text{T}} \\ Q_1 & & \\ Q_2 & & O_{2MN} \end{bmatrix}}_{A} \underbrace{\begin{bmatrix} \hat{\boldsymbol{v}} \\ \hat{\boldsymbol{\lambda}}_1 \\ \hat{\boldsymbol{\lambda}}_2 \end{bmatrix}}_{x} = \underbrace{\begin{bmatrix} 2 \, Q_0^{\text{T}} \left( \boldsymbol{\omega}^{\circ \frac{1}{2}} \circ \boldsymbol{\delta} \circ \boldsymbol{v}_D \right) \\ \mathbb{O}_{MN} \\ Q_2 \boldsymbol{v}_w \end{bmatrix}}_{b}.$$

(6)

The $\boldsymbol{A}$ matrix is real, sparse, symmetric, and of size ($4MN \times 4MN$). The column vector $\boldsymbol{v}_D$ represents the (negative) echocardiographic Doppler velocities, which are commonly calculated by a one-lag autocorrelator of I/Q ultrasound signals (Madiena et al., 2018). The column vector $\boldsymbol{v}_w$ includes the radial and angular components of the endocardial velocities, which can be estimated by speckle tracking (Garcia et al., 2017) or deep learning (Evain et al., 2020), for example.

| | matrices size = $(M \times N)$ unless specified | column vectors length = $(MN)$ unless specified | description |
|---|---|---|---|
| **input data** | $\boldsymbol{V_D}$ | $\boldsymbol{v_D}$ | Negative Doppler velocities before scan conversion. |
| | $\boldsymbol{R}$ | $\boldsymbol{r}$ | Radial coordinates of the grid nodes. |
| | $\boldsymbol{W}$ | $\boldsymbol{\omega}$ | Weights allocated to the Doppler data (*in vivo* only) |
| | $\boldsymbol{V_{Wr}}$ | $\boldsymbol{v_{Wr}}$ | Radial components of the endocardial wall velocities. |
| | $\boldsymbol{V_{W\theta}}$ | $\boldsymbol{v_{W\theta}}$ | Angular components of the endocardial wall velocities. |
| | $\boldsymbol{N_r}$ | $\boldsymbol{n_r}$ | Radial components of the unit vector normal to the cardiac inner wall. Is zero if the node does not belong to the endocardium. |
| | $\boldsymbol{N_\theta}$ | $\boldsymbol{n_\theta}$ | Angular components of the unit vector normal to the cardiac inner wall. Is zero if the node does not belong to the endocardium. |
| | $\boldsymbol{\Delta}$ | $\boldsymbol{\delta}$ | Binary array that defines the left ventricular cavity. It is 1 if the node is inside or on the edge of the left ventricular cavity, 0 otherwise. |
| **output** | $\boldsymbol{\hat{V}_r}$ | $\boldsymbol{\hat{v}_r}$ | Radial components of the estimated velocities. |
| | $\boldsymbol{\hat{V}_\theta}$ | $\boldsymbol{\hat{v}_\theta}$ | Angular velocities to be estimated |
| | | $\boldsymbol{\hat{v}}$ | Column vector of length $(2MN)$ that contains the estimated velocities. It is part of the solution of the constrained minimization problem. $$\hat{v} = [\hat{v}_r^{\mathrm{T}} \ \hat{v}_\theta^{\mathrm{T}}]^{\mathrm{T}}$$ |
| | | $\boldsymbol{\lambda_1}$ | Lagrange multipliers for the 1st constraint (divergence-free) |
| | | $\boldsymbol{\lambda_2}$ | Lagrange multipliers for the 2nd constraint (free-slip boundary conditions) |
| **other arrays** | $\boldsymbol{I_q}$ | | $\boldsymbol{I_q}$ is the identity matrix of size $(q \times q)$. |
| | $\boldsymbol{O_q}$ | $\mathbb{O}_q$ | $\boldsymbol{O_q}$ is the null matrix of size $(q \times q)$. $\mathbb{O}_q$ is a column vector of zeros of size $(q \times 1)$, where $q$ is a general length. |
| | $\boldsymbol{\dot{D}_q}$ | | First-order derivative operator matrix of size $(q \times q)$ based on a second-order central finite difference (see appendix). |
| | $\boldsymbol{\ddot{D}_q}$ | | Second-order derivative operator matrix of size $(q \times q)$ based on a second-order central finite difference (see appendix). |

**Table 1.** Column arrays and matrices used in the linear system that describes the constrained minimization problem.

Since $\boldsymbol{v_D}$ and $\boldsymbol{v_w}$ can be significantly noisy, so can be the estimated velocity vector field $\hat{\boldsymbol{v}}$ in the solution $\boldsymbol{x}$ of the linear system (6). We thus added a smoothing regularizer $\mathcal{S}$ and solved (6) using a regularized least-squares approximation:

$$\hat{\boldsymbol{x}} = \arg\min_{\boldsymbol{x}} \{\|\boldsymbol{Ax} - \boldsymbol{b}\|^2 + \alpha\|\mathcal{S}(\boldsymbol{x})\|^2\}. \tag{7}$$

To ensure spatial smoothing in both radial and angular directions, as in (Assi et al., 2017), we defined $\mathcal{S}$ by

$$\mathcal{S}(v_r, v_\theta) = \sum_{m \in \{r, \theta\}} \left\{ \left(r^2 \frac{\partial^2 v_m}{\partial r^2}\right)^2 + 2\left(r\frac{\partial^2 v_m}{\partial r \partial \theta}\right)^2 + \left(\frac{\partial^2 v_m}{\partial \theta^2}\right)^2 \right\}. \tag{8}$$

The scalar $\alpha > 0$ is the regularizing parameter. It must be chosen to provide a good trade-off between under- and over-fitting. As explained in the next paragraph, $\alpha$ was determined by analyzing the $L$-curve (Hansen and O'Leary, 1993; Hansen, 2000). The regularized least-squares problem (7) can be written as

$$\hat{\boldsymbol{x}} = \arg\min_{\boldsymbol{x}} \{\|\boldsymbol{Ax} - \boldsymbol{b}\|^2 + \alpha\|\boldsymbol{Sx}\|^2\}, \text{ with } \boldsymbol{S} = [1\ 0] \otimes \boldsymbol{I_2} \otimes \boldsymbol{Q_3}, \tag{9}$$

where $\boldsymbol{Q_3}$ is the matrix of size ($6MN \times 2MN$) defined by (Assi et al., 2017):

$$\boldsymbol{Q_3} = \begin{bmatrix} \frac{1}{h_r{}^2}\left(\operatorname{diag}(\boldsymbol{\delta} \circ \boldsymbol{r}^{\circ 2}) \left(\boldsymbol{I_N} \otimes \boldsymbol{\ddot{D}_M}\right)\right) \\ \frac{\sqrt{2}}{h_r h_\theta}\left(\operatorname{diag}(\boldsymbol{\delta} \circ \boldsymbol{r}) \left(\boldsymbol{\dot{D}_N} \otimes \boldsymbol{\dot{D}_M}\right)\right) \\ \frac{1}{h_\theta{}^2}\left(\operatorname{diag}(\boldsymbol{\delta}) \left(\boldsymbol{\ddot{D}_N} \otimes \boldsymbol{I_M}\right)\right) \end{bmatrix}. \tag{10}$$

From (9), the solution $\hat{\boldsymbol{x}}$ finally verifies

$$\underbrace{\left(\boldsymbol{A}^{\mathrm{T}}\boldsymbol{A} + \alpha\,\boldsymbol{S}^{\mathrm{T}}\boldsymbol{S}\right)}_{\boldsymbol{M}} \hat{\boldsymbol{x}} = \boldsymbol{A}^{\mathrm{T}}\boldsymbol{b}. \tag{11}$$

The $\boldsymbol{M}$ matrix is real, sparse, positive semi-definite, and of size ($4MN \times 4MN$). The first $2MN$ elements of the solution vector $\hat{\boldsymbol{x}}$ contains the radial and angular components of the estimated velocities $\hat{\boldsymbol{v}}^{\mathrm{T}} = \left[\hat{\boldsymbol{v}}_r{}^{\mathrm{T}}\ \hat{\boldsymbol{v}}_\theta{}^{\mathrm{T}}\right]$. The $\boldsymbol{M}$ matrix is rank-deficient because it contains columns and rows of zeros, as the region of interest does not cover the entire domain. After having discarded the null rows and columns to make the matrix full-rank and positive-definite, we solved the sparse linear system (11) by using Cholesky decomposition. We have solved the system (11) in MATLAB language. It took about 0.2 s to create the matrices and calculate the solution with a personal laptop (Intel Core i5, 2.5GHz).

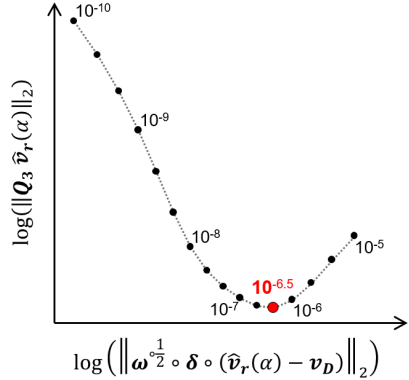

**Figure 3.** Unsupervised selection of the regularization parameter – After fitting the $L$-curve from a set of regularization parameters (black dots), we choose the regularization parameter that achieves the global minimum (here $10^{-6.5}$). This example is from a patient at the end of early filling.

## 2.2. Choice of the regularization parameter.

In contrast with the previous $i$VFM algorithm that contained three regularization parameters, the new physics-constrained version includes a single one ($\alpha > 0$). This strategy simplifies the solution of the problem. The $L$-curve method (Hansen and O'Leary, 1993) is one approach for the selection of a single regularization parameter. It identifies the trade-off between the amount of regularization and the quality of the fit to the given data. The $L$-curve consists of a log-log plot of the residual norm versus the regularization norm for a set of regularization parameter values. The $L$-curve associated with our minimization problem (9) was

$$\left\{\log\left(\left\|\boldsymbol{\omega}^{\circ\frac{1}{2}} \circ \boldsymbol{\delta} \circ (\hat{\boldsymbol{v}}_r - \boldsymbol{v_D})\right\|_2\right), \log(\|\boldsymbol{Q_3}\,\hat{\boldsymbol{v}}_r\|_2)\right\}. \tag{12}$$

An appropriate regularization parameter $\alpha_c$ can be the one that maximizes the curvature of the $L$-curve (Hansen, 2000) or that located at the inflection point (Milovic et al., 2021). We used the former method to determine $\alpha_c$. The $L$-curve method requires solving the system (11) with several values of $\alpha$. For reasons of computational time, it is preferred not to repeat this process for each Doppler image. Therefore, we calculated the $L$-curve and the $\alpha_c$ parameter once, at the end of the early filling, and used this $\alpha_c$ value for the other Doppler fields. We chose the end of the early filling because this is our time of interest, when the vortex forms. We therefore sought to optimize the regularization parameter at this particular time. To estimate $\alpha_c$, we fitted the $L$-curve with a polynomial function for a set of $\alpha$ parameters. We then determined the regularization parameter that maximized the curvature. In our cases, it was also the parameter that reached the global minimum of the $L$-curve (Figure 3), as all $L$-curves were convex in our study.

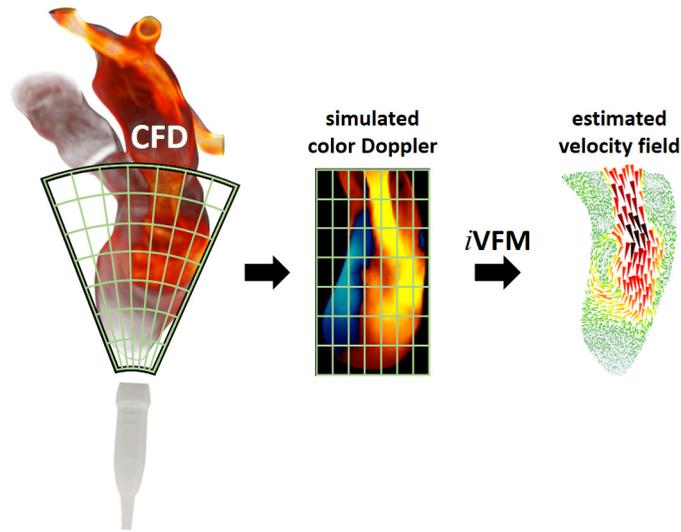

**Figure 4.** From left to right: the physics-constrained *i*VFM algorithm was tested in a patient-specific CFD model of the left heart flow. Color Doppler fields were simulated from the radial velocity components. The velocity vector fields estimated by *i*VFM were compared with the ground-truth CFD fields.

### 2.3. Analysis in a patient-specific CFD heart model.

The physics-constrained *i*VFM was tested under the same conditions as the previous version. We used a patient-specific physiological CFD (computational fluid dynamics) model of cardiac flow developed by Chnafa *et al*. (Chnafa et al., 2014, 2016). The CFD cardiac cavities, as well as their dynamics, were issued from images acquired by computed tomography (Figure 4). Large amplitude motions were treated by adopting an arbitrary Lagrangian-Eulerian (ALE) method. Several cardiac cycles of intracardiac flow were simulated in the left heart. Color Doppler velocities were simulated from the phase-averaged CFD velocities. An apical three-chamber view was reproduced (Figure 4) to obtain a Doppler sector including the apex, mitral inlet, and left ventricular outflow tract. Simulated Doppler images were obtained in a polar (fan-shaped) grid from the radial velocity components (50 scanlines and 160 samples/scanline, which gave angular and radial steps of 0.9 degrees and 0.61 mm). Zero-mean Gaussian white noise with velocity-dependent local variance (Jensen, 1996) was added to obtain signal-to-noise ratios (SNR) ranging between 10 and 50 dB [see equations (10) and (11) in (Muth et al., 2011) and the supplementary document]. We simulated 100 color Doppler images evenly distributed over a cardiac cycle. The radial and angular velocity components were estimated by *i*VFM through solving the linear system (11). No weights were allocated to the simulated Doppler data ($\omega = 1$, everywhere). The regularization parameter $\alpha_c$ was determined (at the end of early filling) by using the $L$-curve (12). We compared the *i*VFM-derived velocity fields with the original

CFD fields. For both the radial and angular components, we calculated the root mean squares errors normalized by the maximum velocity defined by

$$nRMSE = \frac{1}{\max\|\vec{v}_{CFD}\|} \sqrt{\frac{1}{n} \sum_{k=1}^{n} \left\| \vec{v}_{iVFM_k} - \vec{v}_{CFD_k} \right\|^2}. \tag{13}$$

The parameter $n$ stands for the number of velocity samples in the left ventricular cavity. We pooled the radial and angular components of the 100 $i$VFM fields to calculate linear regression coefficients ($i$VFM vs. CFD). From the perspective of being able to characterize the intraventricular flow as a whole, we also calculated two global parameters: the spatial averages of the vorticity and the absolute value of the stream function. The vorticity $\omega$ (in $s^{-1}$) is given by the curl of the vector field. In polar coordinates, it is written as (Yu and Tian, 2013; Mehregan et al., 2014)

$$\omega = \frac{1}{r} \left( \frac{\partial (r v_r)}{\partial r} - \frac{\partial v_\theta}{\partial \theta} \right). \tag{14}$$

The stream function ($\psi$) is defined by the following differential equations (Yu and Tian, 2013)

$$v_r = \frac{1}{r} \frac{\partial \psi}{\partial \theta} \; ; \; v_\theta = -\frac{\partial \psi}{\partial r}. \tag{15}$$

For each frame, the constant was defined so that the integral of $\psi$ over the surface of the left ventricle was zero. In an incompressible 2-D flow, the isolines of a stream function represent the streamlines.

### 2.4. Analysis in patients.

We tested the new physics-constrained $i$VFM in patient data (no valvular regurgitation, no arrhythmia) to illustrate its feasibility in a clinical context. Echo-Doppler images of the left ventricle were acquired in the apical long-axis three-chamber view using a Vivid e95 ultrasound scanner (GE Healthcare) and a 2.9-MHz phased array (M4S). Doppler data were extracted before scan conversion (*i.e.*, in a polar grid whose radial directions are those of the scanlines) using EchoPAC (GE Healthcare). The Doppler velocities were de-aliased using the technique described in (Muth et al., 2011), and the inner left ventricular boundaries were segmented manually. The intraventricular vector flow maps were estimated by $i$VFM. In clinical practice, high Doppler power is generally associated with reliable Doppler velocity. The weights $\omega$ [see Eq. (2)] were then defined from the power Doppler fields. In EchoPAC, power Doppler ($P_D$) is ranged between 1 and 100. We used the following weights:

$$\omega = \log(P_D)/2. \tag{16}$$

We choose the regularization parameter by using the *L*-curve method (see paragraph *Choice of the regularization parameter*) at the end of early filling. The same regularization parameter was used to calculate the other intraventricular vector flow fields of the cardiac cycle.

## 3. Results

### 3.1. Ground-truth vs. iVFM-derived velocities
Figure 5 depicts the early left ventricular filling and vortex formation in the left-heart CFD model, as estimated by $i$VFM from the Doppler velocity components. After pooling all the radial and angular velocities, their coefficients of determination were $r^2 = 0.98$ and $r^2 = 0.63$ respectively (Figure 6). The normalized root mean square errors ranged between $0.3 - 4\%$ and $1.7 - 12\%$ for the radial and angular velocities, respectively (Figure 7).

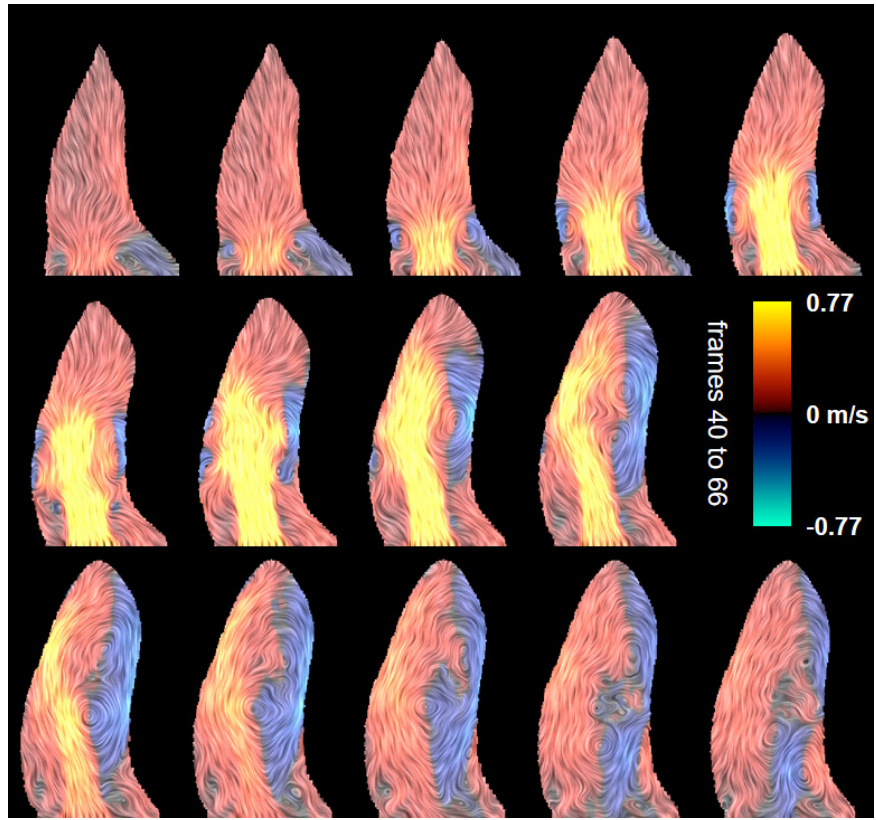

**Figure 5.** Intraventricular flow maps recovered by *i*VFM (in the CFD model) from the Doppler velocities. The LIC (line integral convolution) patterns represent the streamlines. The Doppler velocities are presented in red and blue colors.

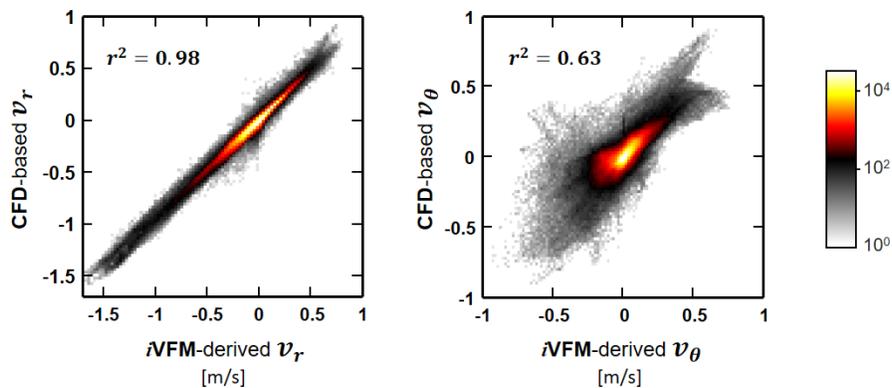

**Figure 6.** CFD-based vs. *i*VFM-derived velocities. Velocity data from the 100 CFD images were pooled. The binned scatterplots display the number of velocity occurrences.

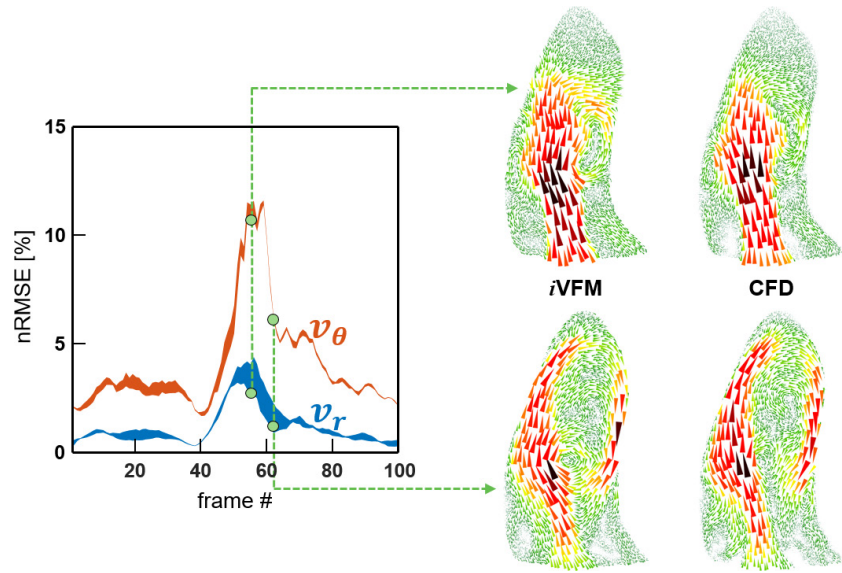

**Figure 7.** Normalized root mean square errors (nRMSE) between the *i*VFM-derived and CFD velocity vectors. The thickness of the curves reflects the range of errors as a function of the Doppler noise (SNR from 10 to 50 dB).

### 3.2. Vorticity and stream function

The mean vorticity (Figure 8) was maximal around frame #60 (2nd snapshot of the last row in Figure 5; see also Figure 9), at the end of the left ventricular relaxation, and reached a peak of ~10 s$^{-1}$. CFD-based and *i*VFM-derived vorticities were concordant ($r^2 = 0.97$), with a difference of $1.7 \times 10^{-3} \pm 1.6 \times 10^{-3}$ s$^{-1}$.

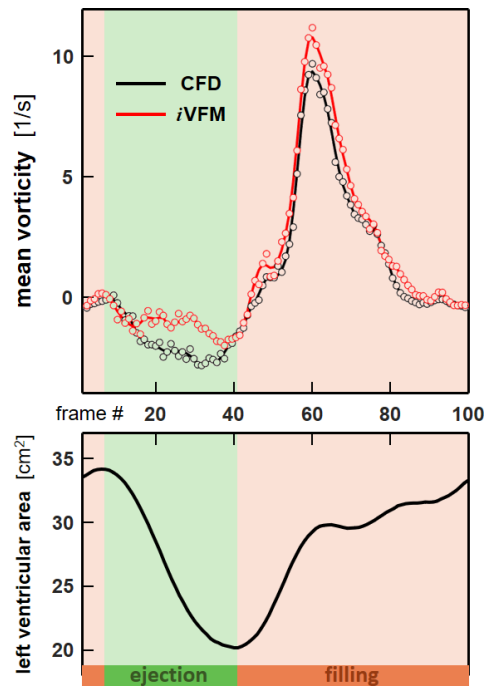

**Figure 8.** CFD-based vs. *i*VFM-derived mean vorticity. The vorticity was averaged over the area of the left ventricular intracavitary cross-section.

A series of stream functions over a cardiac cycle is depicted in Figure 9 to highlight the streamlines. An animation is given in the supplementary document to appreciate the vortex formation during diastole. CFD-based and *i*VFM-derived stream functions were concordant ($r^2 = 0.88$). The mean of their absolute values reached local maxima during ejection and early filling.

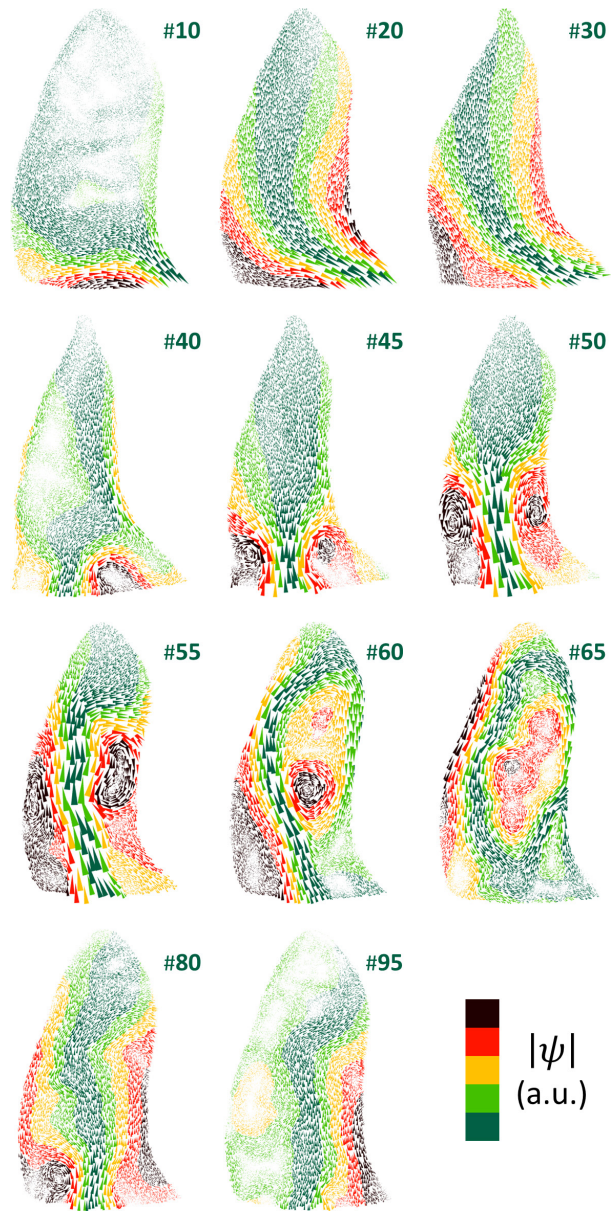

**Figure 9.** Velocity fields and stream functions (their absolute values) over a cardiac cycle. The green numbers refer to the frame numbers (see also Figure 8).

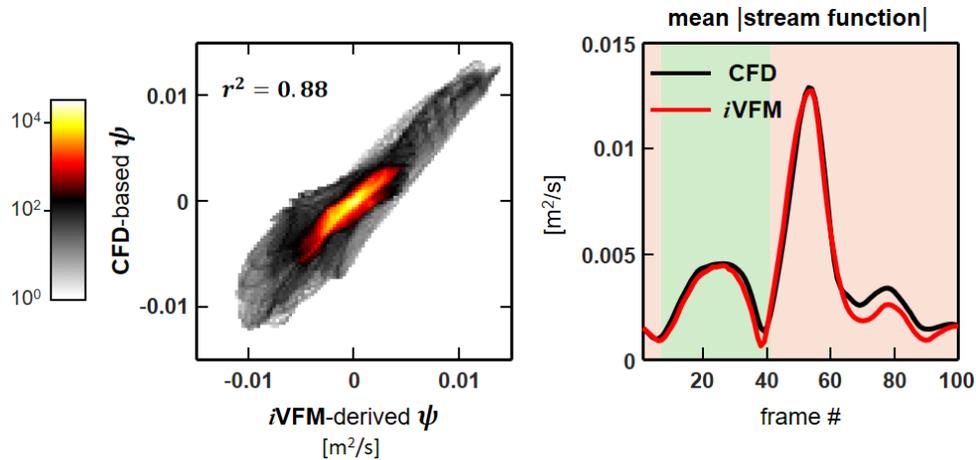

**Figure 10.** Left: CFD-based vs. *i*VFM-derived stream function. Data from the 100 CFD images were pooled. The binned scatterplot (left panel) displays the number of velocity occurrences. Right: Spatial average of the absolute value of the stream function.

### 3.3. Vector flow mapping in a clinical context

The vector flow maps created with the new *i*VFM algorithm highlighted intraventricular flows, otherwise hardly discernible by standard color Doppler. An example of blood flow dynamics during a cardiac cycle is shown in Figure 11 (an animation is given in the supplementary document). This example shows the formation of a large vortex in a normal patient (no heart disease) during early filling (i.e. ventricular relaxation). The vortex was still visible during diastasis, the period between ventricular relaxation and atrial contraction. Figure 12 represents snapshots of intraventricular blood flow during early filling in nine patients. The vortex ring is visible in some images at the beginning of early filling. In others, the large vortex that formed at the end of early filling can be seen.

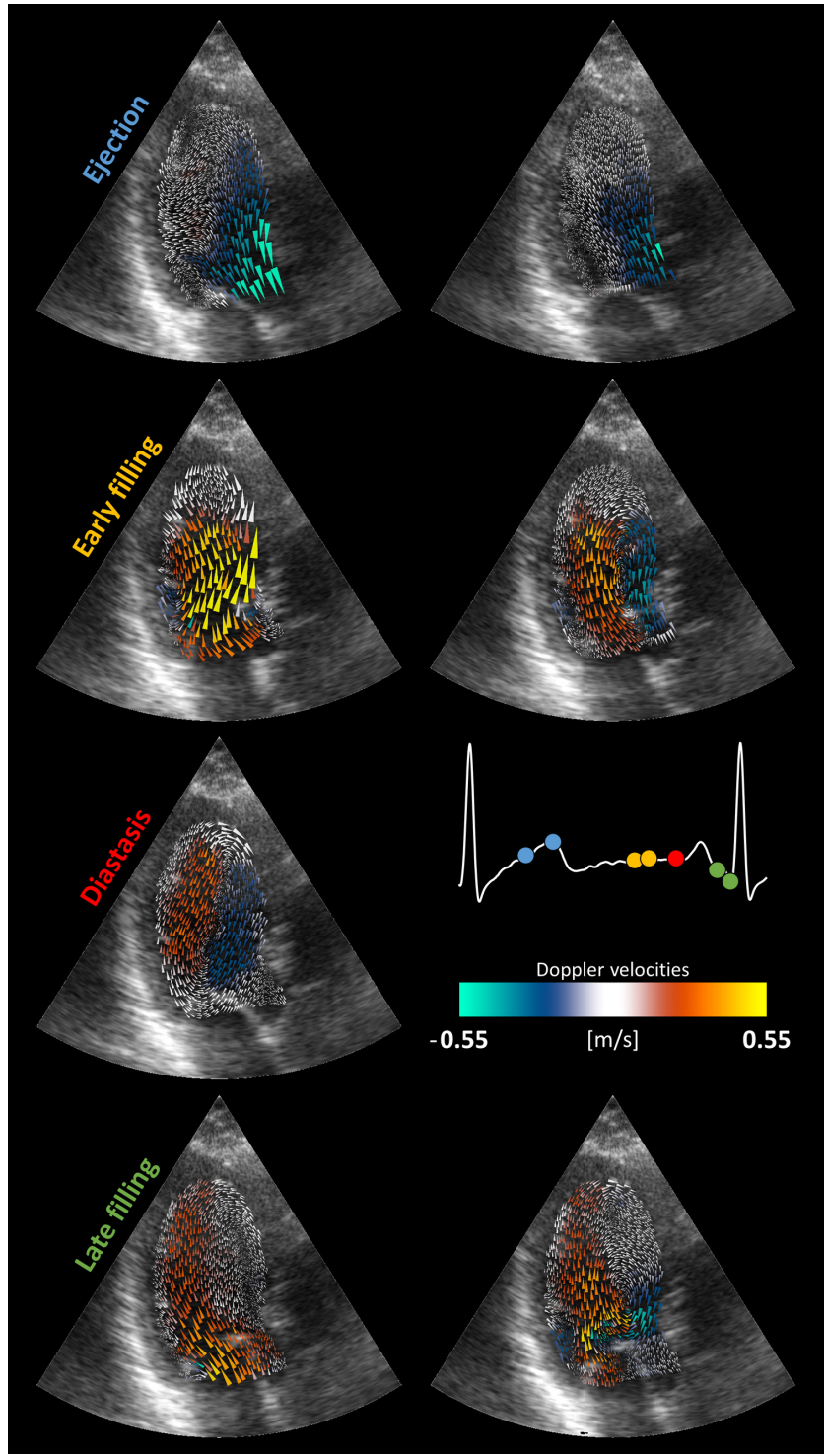

**Figure 11.** Physics-constrained intraventricular vector flow mapping (*i*VFM) in a patient. Selected frames show blood inflow and outflow in the left ventricle. The large vortex that forms during early filling is visible during diastasis. The color of the arrows represents the original color Doppler fields from which the *i*VFM fields were deduced.

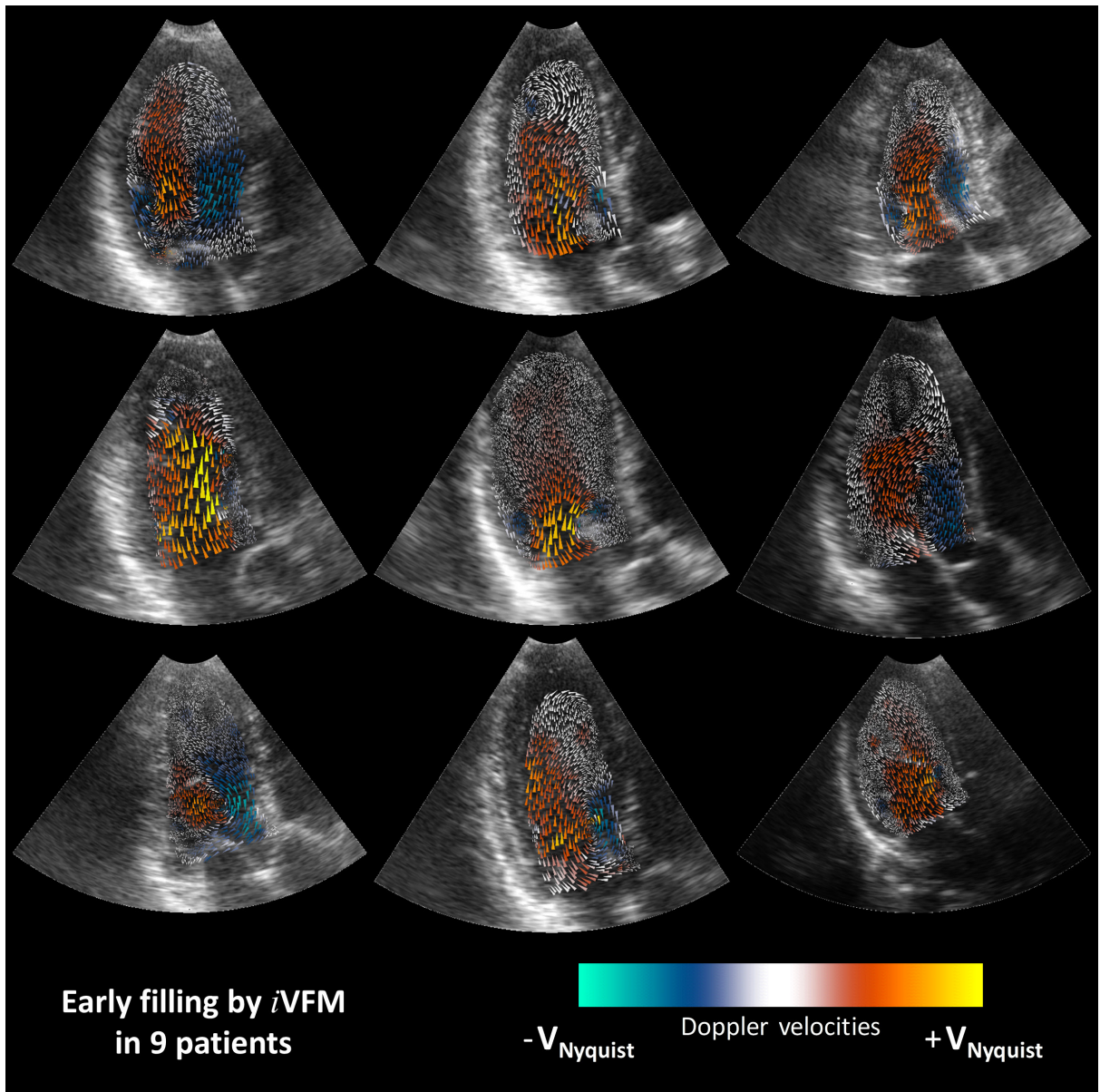

**Figure 12.** Physics-constrained intraventricular vector flow mapping (*i*VFM) in nine patients. These selected frames display intraventricular blood flow during early filling (i.e. ventricular relaxation). The color of the arrows represents the original color Doppler fields from which the *i*VFM fields were deduced.

## 4. Discussion

We have introduced a physics-constrained version of the *i*VFM algorithm for the generation of 2-D intra-ventricular velocity vector fields from Doppler echocardiography. The least-squares regularization method is similar to that described in our previous paper (Assi et al., 2017). However, in contrast to our former work, the free-divergence and boundary conditions are no longer expressed in the least-squares sense but are now set explicitly. The physical constraints reduce the number of regularization parameters to one, instead of three. Using a finite difference scheme and the method of Lagrange multipliers, the minimization problem reduces to a sparse linear symmetric system that can be solved numerically through standard methods. This physics-constrained *i*VFM has formed the framework of a volumetric three-component version (3D-*i*VFM) based on clinical triplane echocardiography. A beta version of 3D-*i*VFM is briefly described in (Vixège et al., 2021).

### 4.1. Limitations of color Doppler and iVFM

Intracardiac blood flow is three-dimensional and unsteady. Any approach to reconstruct the actual velocity field exactly, from single-component data such as provided by color Doppler, is bound to fail. Only an estimated field can be recovered because some hemodynamic information is missing. To obtain an acceptable estimate, one must resort to assumptions supported by physics. In *i*VFM, we assume that the out-of-plane components are negligible in the 3-chamber view. From a physical standpoint, since blood is incompressible under the conditions studied, this means that we assume that the flow is divergence-free on this plane. It is untrue in practice since the actual flow is not planar. The ventricular and valvular geometries, however, induce that the three-chamber plane is a plane of quasi-symmetry in normal subjects (Pedrizzetti and Domenichini, 2005). Accordingly, velocimetry by 4-D phase-contrast cardiac magnetic resonance shows that the intraventricular flow pathlines are essentially symmetric with respect to this plane (Töger et al., 2012; Markl et al., 2011). The *i*VFM method is also limited by the intrinsic spatial resolution of color Doppler. The latter depends on several factors (some of which cannot be controlled by the clinician): 1) center frequency, 2) pulse length, 3) elevation focus, 4) beamforming grid steps, 5) autocorrelator numerical scheme, etc. As an example, the resolution of the Doppler grids of the nine patients in Figure 12 were $0.61 \pm 0.17$ mm by $1.4 \pm 0.37$ degrees. Because the spatial resolution of color Doppler is limited, the boundary layer cannot be measured. It is therefore consistent to rely on free-slip boundary conditions. It should also be noted that color Doppler cannot measure turbulent fluctuations. Indeed, for a given pixel location, the Doppler velocities are generally constructed by an average autocorrelation following ~8 successive ultrasound transmissions emitted at nearly 4000 Hz, which gives a temporal scale of ~2 ms. Furthermore, it is preferred to use a kernel around this pixel to reduce the variance of the Doppler estimator. Intrinsically, color Doppler has low spatiotemporal resolution and is therefore not suitable for measuring turbulent properties. To complicate matters, in cardiac imaging, color Doppler contains significant clutter from stationary or moving myocardial tissue, which requires filtering to mitigate their negative effects. How clutter filtering and the resulting dropouts affect the velocity reconstruction by *i*VFM was not investigated in this work. An approach for more comprehensive analyses would be the use of ultrasound simulations (Garcia, 2021) after seeding the flow with scattering particles (Swillens et al., 2010; Shahriari and Garcia, 2018). Given the limitations of color Doppler, *i*VFM can only provide a velocity field smoothed in time and space. Although several researchers have asserted that energy dissipation due to blood viscosity in turbulent flow can be measured by *i*VFM (Stugaard et al., 2015; Zhong et al., 2016), this claim is incorrect. The main reason is that the kinetic energy of turbulence is dissipated into heat by viscous forces at Kolmogorov scales, which are the smallest scales of turbulent flow. Such spatial and temporal scales cannot be captured by color Doppler.

### 4.2. iVFM's ability to recover large-scale flow patterns

Based on our results in the CFD cardiac model, *i*VFM can accurately catch the global dynamics of the intraventricular flow. The normalized errors ranged from 2% to 12% for the crossbeam (angular) velocity components and were less than 5% during most of the cardiac cycle. The normalized errors were less than

4% for the axial (radial) velocity components. Errors in the radial direction are mainly due to the two constraints that are not entirely satisfied (i.e. incompressible planar flow and free-slip boundary conditions). The concordance of the stream functions ($r^2 = 0.88$, Figure 10) between the actual and estimated flow fields shows that *i*VFM can successfully decipher the large-scale features. The stream function is defined for divergence-free flows in two dimensions and is therefore well suited to the physics-constrained *i*VFM. It is constant along a streamline. The *i*VFM-CFD match provided evidence that the main flow directions were successfully retrieved by *i*VFM from the Doppler components. The *i*VFM algorithm also provided an accurate estimate of the mean intraventricular vorticity (Figure 8). Vorticity reflects the local rate of rotation of a fluid particle. The mean vorticity reached a maximum at the end of early filling, i.e., when the vortex was largest. This peak in mean vorticity could reflect the grade of filling of the left ventricle. This potential biomarker of diastolic function should be tested in patients with heart failure. We hypothesize that it is likely to decline with impaired filling.

### 4.3. iVFM and derived forms

The idea of recovering a planar velocity vector field from color Doppler information was introduced by Ohtsuki and Tanaka (Ohtsuki and Tanaka, 2006). The proposed method was further described by Uejima *et al.* (Uejima et al., 2010), who decomposed the intraventricular flow into a flow called "basic" and an axisymmetric vortex flow. The axisymmetry constraint is not realistic under physiological conditions since the vortex ring stretches and deforms into an elongated shape. The *i*VFM algorithm was introduced by Garcia *et al.* (Garcia et al., 2010). In this version, the 2-D polar continuity equation is integrated perpendicular to the ultrasound scanlines, for a given radial distance from the cardiac phased array. This technique has been implemented in FUJIFILM Healthcare ultrasound scanners (Tanaka et al., 2015). The main limitation is that the integrating operators work isoradially, i.e. the solution on an isoradial line does not depend on the neighboring lines. In a patent, Pedrizzetti and Tonti (Pedrizzetti and Tonti, 2012) broke down the velocity vector field as the sum of the Doppler field and an irrotational (curl-free) velocity field. The curl of the estimated field is therefore equal to that of the color Doppler velocity field, which has no physical or physiological support. Jang *et al.* (Jang et al., 2015) added a source term to the Navier-Stokes equation, then introducing an additional unknown into the system to be solved. Because the problem was ill-posed, the authors sought the minimum-norm solution, which has little sense from a physical and physiological perspective. Assi *et al.* (Assi et al., 2017) reformulated the *i*VFM algorithm in 2-D using a regularized least-squares method. The divergence-free and boundary conditions were written in the least-squares sense. With an additional second-order smoother, this resulted in three regularization parameters that were determined through an *L*-hypercurve. This numerical limitation is solved with the version described in the present paper, which requires only one regularization parameter. Compared to Assi's version, the reconstruction errors were alike (see the supplementary document). The first and second versions of *i*VFM were investigated in the context of high-frame-rate (ultrafast) echocardiography by (Yu et al., 2017) and (Faurie et al., 2017), respectively. Meyers *et al.* (Meyers et al., 2020) reconstructed the velocity vector field using a Laplace equation that relates the streamfunction and the vorticity. This formulation is also based on a 2-D divergence-free assumption, which makes it close to *i*VFM. As inlet (mitral) flow conditions, the authors predefined a velocity profile whose amplitude was given by pulsed-wave Doppler. While this seems like a wise option, this strategy burdens the method with additional processing. The results obtained were close to those of *i*VFM in the apical long-axis view.

### 4.4. Improvements to the latest version of the iVFM

The physics-constrained *i*VFM described in this work is an improved version of the previous one. The divergence-free and boundary constraints are no longer written in the least-squares sense but are expressed through equality constraints. The problem can be solved using the Lagrange multiplier method. It is important to note that the physics-constrained *i*VFM contains only one regularization parameter (instead of three in the previous version), which greatly simplifies the resolution of the problem and makes it more robust. The two technical limitations that would need to be improved for easy clinical use of *i*VFM are 1) segmentation of the inner wall of the left ventricle (endocardium), 2) elimination of aliasing. 1) In this study, segmentation was performed manually for the analysis of the clinical cases. This allowed us to determine the positions and velocities of the boundaries that are both involved in the *i*VFM algorithm. To avoid this time-consuming task, the clinical version of *i*VFM will include deep learning-assisted segmentation and myocardial tracking, as described in Leclerc *et al.* (Leclerc et al., 2020) and Evain *et al.* (Evain et al., 2020). 2) Using clinical ultrasound scanners, aliasing must be removed in post-processing. We cannot use advanced techniques as we did with research ultrasound scanners (Posada et al., 2016). The dealiasing method we used is as introduced by Muth *et al.* (Muth et al., 2011). It depends on an input variable that sometimes had to be adjusted manually. To make the dealiasing fully automatic, we will also resort to deep learning (Nahas et al., 2020). We will then have a ready-to-use *i*VFM software package for clinical routine purposes. It is our opinion that it is best to focus on a single clinical biomarker based on intraventricular flow to ease potential diagnostic use. For the sake of validation, we here presented two global parameters based on vorticity and streamfunction. Whether these have any diagnostic power remains to be demonstrated in a cohort of patients. With the new version of the *i*VFM, other criteria could be evaluated, such as the size of the vortex, or properties related to its dynamics.

In addition to facilitating the transition to a clinical trial, the new *i*VFM is transferable to 3-D. In this three-dimensional perspective, rather than using volumetric Doppler data, whose spatiotemporal resolutions are still limited, we opted for the triplane Doppler mode. Unlike volume Doppler, triplane acquisition provides three long-axis planes (2-, 3-, and 4-chamber views). To create 3-D *i*VFM, we rewrote the minimization problem (2) with the three velocity components in a spherical coordinate system. Although two components are unknown (the polar and azimuthal components), the measured triplane Doppler information might be sufficient to reconstruct an acceptable 3-D intraventricular flow. This seems to be confirmed by our first results on 3-D *i*VFM (Vixège et al., 2021).

## 5. Conclusion

We have introduced and validated a physics-constrained *i*VFM algorithm for intraventricular vector flow mapping using color Doppler echocardiography. This algorithm will form the basis of a turnkey *i*VFM clinical software package. It will allow us to test whether intraventricular vortex analysis can improve the assessment of diastolic function in selected patients with heart failure.

## 6. Acknowledgments

This work was carried out in connection with the LABEX CELYA (ANR-10-LABX-0060) of Université de Lyon, within the program "Investissements d'Avenir" (ANR-11-IDEX-0007). Damien Garcia and Franck Nicoud were funded by the French National Research Agency (ANR) through the "4D-iVFM" project (ANR-21-CE19-0034-01).

Additional information for

# Physics-constrained intraventricular vector flow mapping by color Doppler

Florian Vixège, Alain Berod, Yunyun Sun, Simon Mendez, Olivier Bernard,
Nicolas Ducros, Pierre-Yves Courand, Franck Nicoud, and Damien Garcia

E-mails: Garcia.Damien@gmail.com; Damien.Garcia@inserm.fr; Florian.Vixege@creatis.insa-lyon.fr

This supplemental document provides additional information in response to questions from the reviewers. We thank them for their comments. Each paragraph answers a question that clarifies the *i*VFM method and the methodology that we used in this study.

- **Would the accuracy change for parasternal images where the flow is primarily lateral**?

Doppler information is very limited. So we must have as much of this information as possible. The direction of the intraventricular flow occurring mostly in the long axis, the apical view must be preferred. A parasternal view would require additional boundary conditions. For example, Meyers *et al*. (Meyers, B.A., Goergen, C.J., Segers, P., Vlachos, P.P., 2020. *Colour-Doppler echocardiography flow field velocity reconstruction using a streamfunction–vorticity formulation*. J. R. Soc. Interface 17, 20200741) used PWD-derived mitral and aortic valve velocities to add boundary conditions. This approach requires the data from multiple measurements to be registered and may thus complicate clinical application.

The following figure shows a flow reconstruction in the parasternal view, <u>without</u> using additional mitral-inlet constraints. A significant part of the flow could not be recovered.

**Parasternal by *i*VFM**                    **Reference (CFD)**

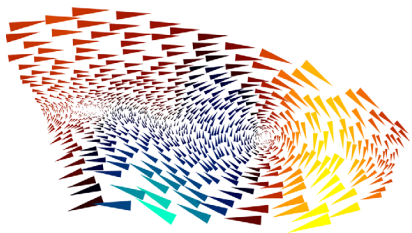
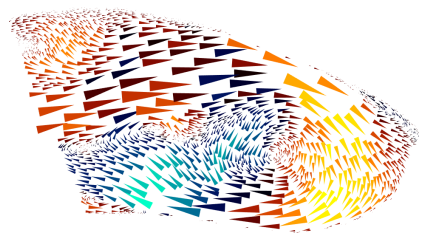

- **How does the present *i*VFM version compare with the previous one**?

The main advantage of the version introduced in this paper is the reduction of the number of regularization parameters from three to one. This makes the system more robust from a numerical point of view, and its resolution faster. We also think (at least, we hope) that it makes the physics of the problem easier to understand. Finally, this approach makes possible the 3-D generalization from triplane color Doppler (Vixège et al., 2021).

The figure below shows the comparison with the previous method that includes three regularization parameters. These three parameters were optimized at the end of the early filling (to yield the smallest errors) and used for the other frames of the cardiac cycle. The errors obtained with the previous *i*VFM (with 40dB-SNR) are represented by the green dotted line.

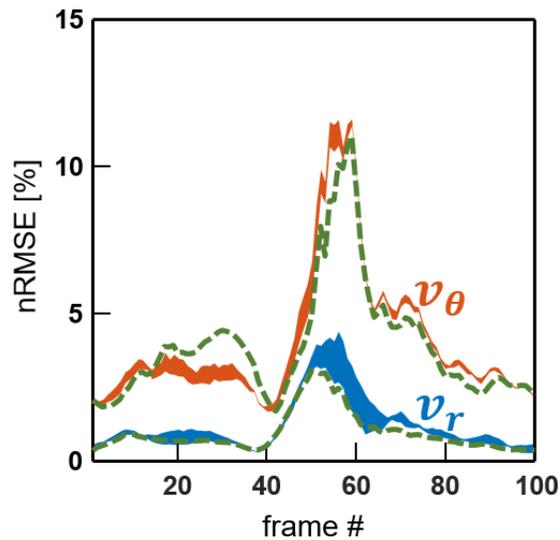

- **What did the noisy CFD-based Doppler velocities look like**?

The Doppler velocities images were simulated from the radial velocity components of the CFD model. We added zero-mean Gaussian white noises with velocity-dependent local variance (Jensen, 1996) to obtain signal-to-noise ratios (SNR) ranging between 10 and 50 dB [see equations (10) and (11) in (Muth et al., 2011)]. The noisy to noise-free Doppler velocities, in a Cartesian coordinate system, looked like:

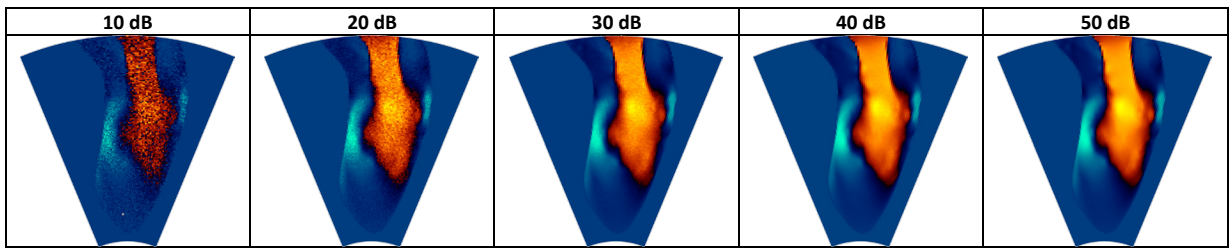

- **<u>Was the solution 2-D divergence-free</u>**? **<u>How was the actual (CFD) 2-D divergence</u>**?

The smoothing term in the *i*VFM induces that the *i*VFM solution is not entirely divergence-free on the three-chamber plane. However, it nearly is, with round-off errors, as can be seen on the histogram below (right panel).

Out-of-plane flows make the 2-D free-divergence assumption incorrect (left panel). The three-chamber plane, however, is a plane of quasi-symmetry, as explained in the main text.

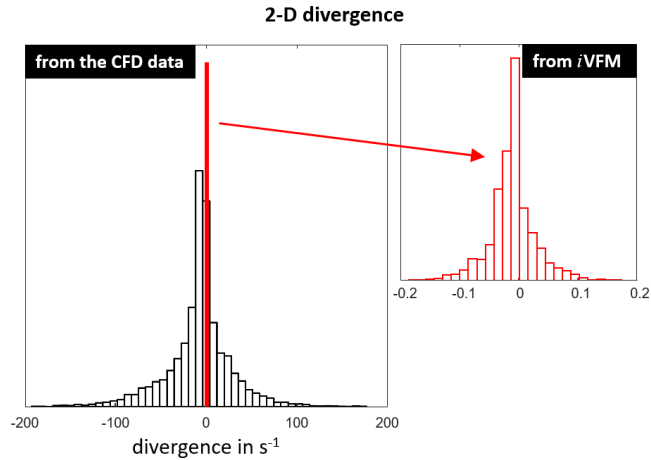

- **<u>What do Doppler velocities typically look like in a patient</u>**?

Doppler data were extracted before scan conversion (i.e., in a polar grid whose radial directions are those of the scanlines) using EchoPAC (GE Healthcare). We recall that we used power Doppler to define a weight function since high Doppler power is generally associated with reliable Doppler velocity.

The following figures show Doppler velocities (size = 177×45) and power-derived weights in one patient, during early filling. They have been resized to a square for clarity.

**<u>Doppler input data</u>**

**<u>Power-derived weights</u>**

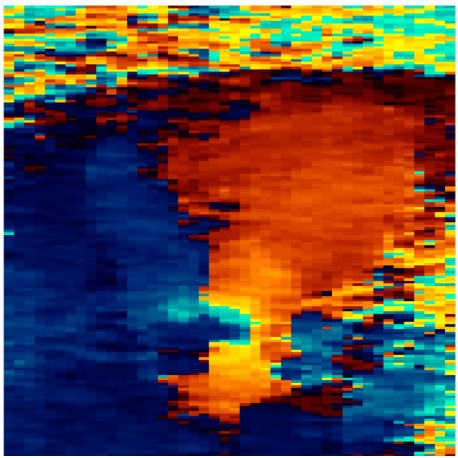 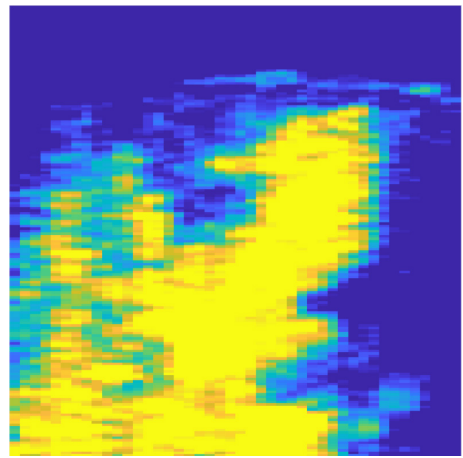